# Corruption and Wealth: Unveiling a national prosperity syndrome in Europe

Juan C. Correa[1, 2, *] and Klaus Jaffe[3]

[1] Departamento de Ciencia y Tecnología del Comportamiento, Universidad Simón Bolívar, Caracas, Venezuela
[2] Facultad de Psicología, Fundación Universitaria Konrad Lorenz, Bogotá, Colombia
[3] Centro de Estudios Estratégicos, Universidad Simón Bolívar, Caracas, Venezuela

* juancorrea@usb.ve

**Abstract**
Data mining revealed a cluster of economic, psychological, social and cultural indicators that in combination predicted corruption and wealth of European nations. This prosperity syndrome of self-reliant citizens, efficient division of labor, a sophisticated scientific community, and respect for the law, was clearly distinct from that of poor countries that had a diffuse relationship between high corruption perception, low GDP/capita, high social inequality, low scientific development, reliance on family and friends, and languages with many words for guilt. This suggests that there are many ways for a nation to be poor, but few ones to become rich, supporting the existence of synergistic interactions between the components in the prosperity syndrome favoring economic growth. No single feature was responsible for national prosperity. Focusing on synergies rather than on single features should improve our understanding of the transition from poverty and corruption to prosperity in European nations and elsewhere.

INTRODUCTION

During the past decade public interest in corruption has grown [1–4]. Roughly speaking, corruption occurs when public officials unlawfully enrich their social network as well as themselves by misusing the power entrusted to them [5]. This phenomenon can be studied by the perception of citizens who participate in household surveys [6]. An important part of the literature has focused on the relationship between corruption and a variety of economic, social and cultural indicators. For instance, it is known that: i) corruption and the economic wealth of a nation are correlated [4–6]; ii) public and private investment are affected by the perception of corruption [7–17]; iii) corruption affects economic growth [18–21], and it is related to income inequality [22] and trade [23–27]; iv) taxation levels are related to economic performance and corruption [8]; v) corruption and inflation interact [28]; vi) nations with high levels of corruption have citizens with characteristic personality traits [29,30]; vi) corruption relates to weak institutions [31]; vii) is linked with pollution [32]; viii) many aspects of societies, such as psychological characteristics of its citizens, and its feelings of well-being [33] are related to corruption perception; ix) corruption is correlated with the physical and human capital of a nation [31]; x) and with important cultural aspects [28,34–39]. Due to the large number of inter-correlated variables that are associated with corruption, a corruption syndrome has been proposed [40–42] that would explain why many countries are poor and have difficulties in engaging in socio-economic development that would allow their citizens to become wealthy.

The idea behind this approach is that national welfare is affected by corruption, not only indirectly through GDP, but also directly through non-material factors like the time and effort required to cope with corrupt behavior, or psychological costs associated with a general climate of unlawfulness [33]. Diener, Diener and Diener [43] found evidence that high income, individualism, human rights, and societal equality correlated strongly with each other and with subjective well-being across surveys; but only individualism persistently correlated with well-being when other predictors were controlled. Corruption and poverty are treated as pathological syndromes that need to be cured [40–42]. Specifically, corruption has often been pinpointed as the major single factor hindering economic progress as it affects economic freedom, socio-political stability and tradition of law abidance.

The socio-cultural characteristics of nations are related with their economic performances. For instance, Chen [44] observed that countries with languages that do not require the future events to be grammatically marked when making predictions (like German) save on average 6% more of their GDP per year and this result is unaffected by the addition of life-cycle-savings controls, it holds in every major region of the world, and appears stable across time. Jaffe and colleagues [45] found that countries with languages that use many synonyms for "guilt" (i.e., "guilt societies") are also countries with high levels of corruption, low governance, difficulties in doing business, low life expectancy and low income per capita. There is no doubt that these and other socio-cultural indicators are linked with the economic performance of countries. Therefore, it is imperative for economists and policy makers to understand the relative importance regarding its effect on corruption, of cultural, psychological and other key indicators and/or predictors of a nation's economic wealth.

There is an important consensus in the literature that economic prosperity is related to trust [31], property rights and the rule of law [46]. These elements in turn are strongly related to the economic complexity [47] and scientific development of a nation [48]. These last two elements, especially the last one, are easy to measure and showed to be highly significant and reliable statistical predictors of the working of national institutions [49]. These studies suggest that an important general predictor of economic wealth is the rationality of a country which is also reflected in the proper working of its institutions. For example, Jaffe et al [48] reported strong correlations between GDP per capita, religious tolerance, and scientific development as estimated by the scientific productivity per capita, whereas in another paper [50] strong correlations between the rule of law, the human development index, scientific productivity, ease of doing business, economic freedoms, transparency and wealth, among other indices, were evidenced. Welsch [33] estimated the total amount of scientists and engineers per population and detected statistically significant relationships between the rationality of countries and the subjective and economic well-being of its inhabitants. Another approach was developed by Hidalgo and Hausmann [49] who estimated the amount of practical knowledge in a society by using the diversity and complexity measures of the products a country exports. These metrics of economic complexity correlated strongly with economic wealth. Studies inspired by this calculus of economic complexity proposed alternative indicators that capture this rationality even better [51]. Employing the scientific productivity of countries, Jaffe et al. [51] developed an indicator called "S-Share", which considers the relative research effort of each scientific subject area as the percentage of the total number of publications of a country, published in journals of that area in a year. This indicator provides robust correlations between scientific productivity in basic sciences with economic growth during the following five years in middle income countries. These studies suggest that scientific development affects multiple aspects of the economy, in such a way that "rationally-governed societies" achieve more wealth for their citizens, and more wealth, in turn, allows for more and better investment in education, science and economic development, increasing wealth further. Yet, several countries have not engaged in this virtuous economic cycle, which suggests that a possible reason for these differences could be related to the predominant psychological, social and cultural idiosyncrasies. If so, existing data related to these characteristics could make it possible to observe human drives that are related with the corruption and wealth of a state.

Clearly then, economic wealth and corruption are manifest covariates. We know that there is a relationship between wealth, measured as GDP per capita, scientific-technological developments and institutions, estimated through the scientific production, measured as publications per capita, and corruption [50]. Here we want to untangle a bit more of this relationship by incorporating into the analysis other psycho-socio-cultural indicators that affect it. To simplify this very complex problem, we restrict our analysis to Europe, where more reliable data is available, economies and cultures are diverse but not too much, and history has fomented high level of information exchange between countries.

METHODS

We built a database composed of several indicators of the European countries. These indicators capture the behavior of nations at different scales ranging from the individual and the societal to the cultural and the economic level. We initially considered the six rounds of the "European Social Survey" (ESS) between 2002 and 2012, both included (http://www.europeansocialsurvey.org/). The ESS is an academically driven cross-national survey that has been conducted every two years across Europe (participant countries in this survey are listed in table A1 in the appendix). The survey offers the possibility to analyze the stability and change in social structure, conditions and attitudes in Europe, allowing inferences on changes in Europe's social, political and moral fabric based on citizens' perceptions and judgments of key aspects of their societies.

We specifically focus on the analysis of the human values that were assessed in the ESS through a modified version of the "Portrait Values Questionnaire" [52]. This questionnaire includes a series of short verbal portraits of different people and each portrait describes a person's goals, aspirations, or wishes that point implicitly to the importance of a single basic value. For instance, "*Thinking up new ideas and being creative is important to him. He likes to do things in his own original way*" describes a person for whom self-direction values are important. "*It is important to him to be rich. He wants to have a lot of money and expensive things*" describes a person who cherishes power values. By describing each person in terms of what is important to him or her, the verbal portraits capture the person's values without explicitly identifying values as the topic of investigation. For each portrait, respondents answer: "*How much like you is this person?*" The response alternatives are; "*very much like me*", "*like me*", "*somewhat like me*", "*a little like me*", "*not like me*", and "*not like me at all*". For each portrait, respondents choose their response by checking one of six boxes labeled with the response alternatives. Thus, respondents' own values are inferred from their self-reported similarity to people who are described in terms of particular values. The similarity judgments are transformed into a 6-point numerical scale. This set of values is listed in Table A4 in the appendix. Along with this set of human values, we also considered the responses to the question of "*How happy are you?*" included in the ESS as a proxy of subjective well-being of citizens.

The original databases for these rounds have at least 40,000 individual responses per round. In order to have a single tractable database, we followed the "macro approach" used by Welsch [33] who analyzed the aggregate (average) response per country. This approach is similar to that employed by the World Bank in the cross-country analysis of the Governance indicators (see also [53]) and proved to be justified because these human values are seen stable across time and within each society with few exceptions [52]. Thus, our analysis consisted in the national average responses given to the human values in each round of the ESS.

In addition, we considered the "corruption perception index", as measured by "Transparency International" (http://www.transparency.org/) for each participant state in each round of the ESS. We reversed the original records of this index to re-adapt its statistical range from 0.1 ("very clean country") to 9.6 ("very corrupt nation"). In our database we also included the annual GDP per capita, the population size of nations and the taxes paid to the central government as percentage of GDP from the "World Development Indicators" of the World Bank (http://databank.worldbank.org/data/databases.aspx). We obtained the total academic productivity of each nation from SCImago (http://www.scimagojr.com/), and divided that number by its total population as given by World Bank, in order to obtain publications per capita. We also obtained the "Human Inequality Index" for 2012 from the online available data provided by the "United Nations Development Program" (https://data.undp.org/).

A conspicuous set of cultural values was assessed through the language spoken in the country. By using Google Translate we quantified the number of synonyms each language provides for "guilt" as the best indicator for separating guilt-shame societies [45]. The resulting database was finally composed by a hundred socio-cultural-economic indicators of the 36 countries that participated in at least one of the six rounds of the ESS.

RESULTS

Table 1 shows the Spearman correlations between corruption perception, wealth measured as GDP per capita, and scientific productivity, in European countries and in the world. The correlation between corruption perception and wealth (GDP per capita) was -0.905 in Europe. The Spearman correlation between corruption perception and publications per capita for European countries was very significant although slightly less strong than the former correlation. The correlation between wealth and publication per capita was also very strong in Europe. Interestingly this correlation using data for all countries in the world with populations above 5 million was even slightly stronger (0.897). That is, wealth and publications per capita was the strongest correlation found among countries worldwide, whereas in Europe the strongest correlation found was between wealth and corruption perception.

**Table 1: Non-parametric correlations of corruption perception, wealth and academic publications per capita in Europe and in the whole World.**

|  | Corruption Perception | | GDP per capita | |
|---|---|---|---|---|
| Data for | Europe | World | Europe | World |
| GDP per capita | -0.905*** | -0.763*** | 1 | 1 |
| Publications per capita | -0.889*** | -0.801*** | 0.887*** | 0.897*** |

NOTE: The correlations included in this table are estimated through Spearman's Rho non-parametric correlation coefficient. Original data of GDP per capita and corruption perception are available in the web pages of The World Bank and Transparency International. The data for publications per capita was estimated using the available data of academic productivity provided by SCImago and divided by the population of each nation as provided by The World Bank. * Correlation is significant at $p < 0.05$; ** significant at $p < 0.01$; *** significant at $p < 0.001$

The Spearman correlation between corruption perception in Europe and the Human Inequality Index of the UNDP was also highly statistically significant: 0.588 ($p < 0.0001$). The correlation between Corruption Perception and the % of GDP paid in taxes was not ($p = 0.14$)

Figure 1 depicts the association between wealth (GDP per capita) and perceived corruption in Europe. The relationship between both variables is practically linear. Two countries, Russia and Norway, deviated somewhat from the linear regression, having higher perceptions of corruption than expected from their levels of wealth. Both are largely dependent on oil exports. The rest of the countries adjust neatly to the inverse relation between corruption perception and wealth: poor countries are perceived as more corrupt than rich ones.

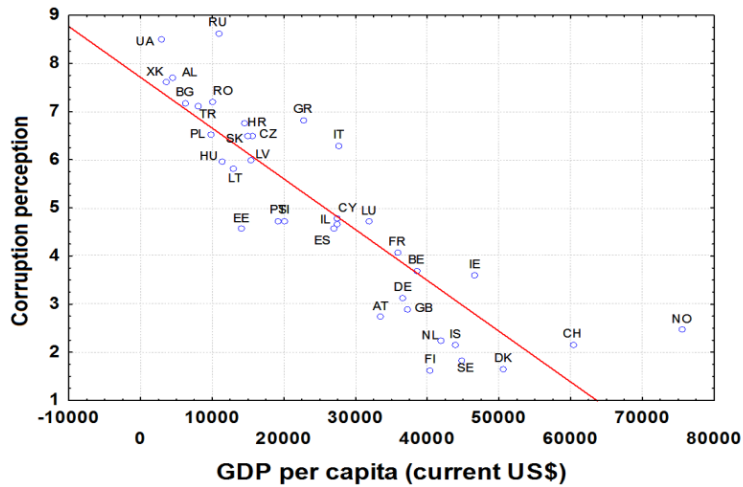

**Figure 1. Association between economic wealth and perceived corruption in 36 European countries.** This figure depicts the scatterplot of the relationship between economic wealth (captured as GDP per capita provided by The World Bank) and the corruption perception (captured as the corruption perception index provided by Transparency International) for the participating countries of the "European Social Survey" (ESS).

Table 2 presents two stepwise regression models explaining corruption perception. The set of human values listed in Model 3 works as important predictors of perceived corruption in European countries ($R^2 = 0.951$; $p < 0.001$), but only a small set of them proved to be statistically significant when GDP per capita is included as a predictor in Model 2 ($R^2 = 0.908$; $p < 0.001$). GDP per capita alone (Model 1) explains a large amount of the variance but statistical significance was somewhat lower than that calculated with non-parametric tests such as those used in Table I. Table II shows the relationships between the prevailing individual values of the citizens in the country, their wealth, and perceived corruption are intimately entangled and statistically not different.

**Table 2: Multivariate relationship of corruption perception, economic wealth and human values**

| Best predictors | Model 1 | Model 2 | Model 3 |
|---|---|---|---|
| GDP per capita (current US$) | 0.87*** (0.08) | -0.43*** (0.09) | |
| Strong government that ensures safety | | -0.36*** (0.09) | -0.25*** (0.11) |
| To make own decisions and be free | | 0.31*** (0.07) | 0.38*** (0.09) |
| To try new and different things in life | | | -0.23*** (0.08) |
| To get respect from others | | -0.27*** (0.07) | -0.41*** (0.07) |
| To be rich, have money and expensive things | | | -0.26*** (0.08) |
| To think new ideas and being creative | | | 0.35*** (0.08) |
| To seek adventures and have an exciting life | | | 0.24*** (0.05) |
| To be loyal to friends and devote to people close | | | -0.30*** (0.09) |
| R-Squared | 0.77*** | 0.91*** | 0.95*** |

NOTE: Each column reports the beta coefficients from stepwise multiple regression models along with its standard errors reported in parentheses. The multivariate association R-squared for each model is presented at the bottom of the table.
* Correlation is significant at $p < 0.05$; ** significant at $p < 0.01$; *** significant at $p < 0.001$

Yet, relationships between the variables studied are strongly non-linear. Thus, non-parametric statistics might be better in unveiling interesting relationships. In Table 3 we present Spearman's correlation between individuals' values surveyed in the ESS and perceived corruption, GDP per capita, synonyms of guilt, publications per capita and the indicator of individual subjective happiness. The values which came out as most statistically significantly correlated were "To follow traditions and customs" and "To have a good time". These, and other questions mainly related to giving importance to individual achievements were negatively correlated to perceived corruption.

The most correlated question "To be rich, have money and expensive things" was classified by the ESS survey as qualifying values related to power. But clearly, monetary achievements are also part of values related to individual achievements. In contrast, answers to questions that correlated positively with corruption perception ere related to values in the domain of benevolence, hedonism and self-direction. That is, countries whose citizens valued strong individual achievements were less corrupt than countries which gave lower importance to these values. Most values that were more important among wealthy countries compared to poorer ones, correlated also with happiness (Table 3). The strongest correlations found were between individual values and the absence of corruption. Individual values correlated only weakly with the presence of corruption. Thus, it is the personality characteristics of citizens living in low corruption countries that provide the statistical significance in the relationship between values and corruption.

**Table 3: Association between individuals' values and perceived corruption, synonyms of guilt, publications per capita, wealth, and happiness for participant countries in the ESS**

| Individual's value | Value Dimension | Perceived corruption | Guilt synonyms | Publications per capita | GDP per capita (current US$) | How happy are you |
|---|---|---|---|---|---|---|
| To have a good time | Hedonism | 0.34* | 0.32 | -0.13 | -0.24 | -0.30 |
| To understand different people | Universalism | 0.27 | 0.31 | -0.49** | -0.35* | -0.45** |
| To be loyal to friends… | Benevolence | 0.25 | 0.38 | -0.26 | -0.27 | -0.41* |
| To seek fun… | Hedonism | 0.22 | 0.33 | -0.24 | -0.29 | -0.31 |
| To think new ideas… | Self-direction | 0.19 | 0.32 | -0.45** | -0.32 | -0.35* |
| To make own decisions and be free | Self-direction | 0.17 | 0.36 | -0.27 | -0.21 | -0.31 |
| To seek adventures… | Stimulation | 0.14 | 0.16 | 0.08 | -0.18 | -0.21 |
| To help people… | Benevolence | 0.04 | 0.23 | -0.25 | -0.15 | -0.27 |
| People are treated equally… | Universalism | 0.00 | 0.13 | -0.33 | -0.12 | -0.26 |
| To try new and different things… | Stimulation | -0.11 | 0.15 | -0.17 | 0.08 | -0.01 |
| To care for nature and environment | Universalism | -0.26 | 0.17 | -0.21 | 0.25 | 0.16 |
| To be humble and modest… | Tradition | -0.40* | -0.09 | -0.16 | 0.30 | 0.25 |
| To do what is told and follow rules | Conformity | -0.54*** | -0.23 | -0.08 | 0.45** | 0.32 |
| Show abilities and be admired | Achievement | -0.58*** | -0.34 | 0.08 | 0.49** | 0.49** |
| To get respect from others | Power | -0.66*** | -0.21 | 0.04 | 0.50** | 0.49** |
| To behave properly | Conformity | -0.67*** | -0.12 | 0.03 | 0.58*** | 0.45** |
| To be successful and… | Achievement | -0.68*** | -0.25 | 0.19 | 0.61*** | 0.60*** |
| To live in secure/safe surroundings | Security | -0.74*** | -0.13 | 0.23 | 0.67*** | 0.60*** |
| Strong government that ensures safety | Security | -0.80*** | -0.21 | 0.19 | 0.72*** | 0.67*** |
| To be rich, have money… | Power | -0.80*** | -0.46* | 0.30 | 0.79*** | 0.75*** |
| To follow traditions… | Tradition | -0.82*** | -0.29 | 0.18 | 0.68*** | 0.65*** |

NOTE: The table presents the correlations between each individual value with perceived corruption, synonyms of guilt, publications per capita, wealth, and happiness for participant countries in the ESS. The correlations are estimated through Spearman's Rho non-parametric correlation coefficient. Individual values are classified in ten conceptual dimensions that summarize its relationships with general human motivations. * Correlation is significant at $p < 0.05$; ** significant at $p < 0.01$; *** significant at $p < 0.001$

Richest citizens living in counties with low corruption perception gave high priorities to the highly negatively correlated values that relate to: tradition, conformity, security, achievement, and power. Figure 2 depicts the association between perceived corruption and importance assigned to the two values surveyed by ESS that predict perceived corruption positively and negatively.

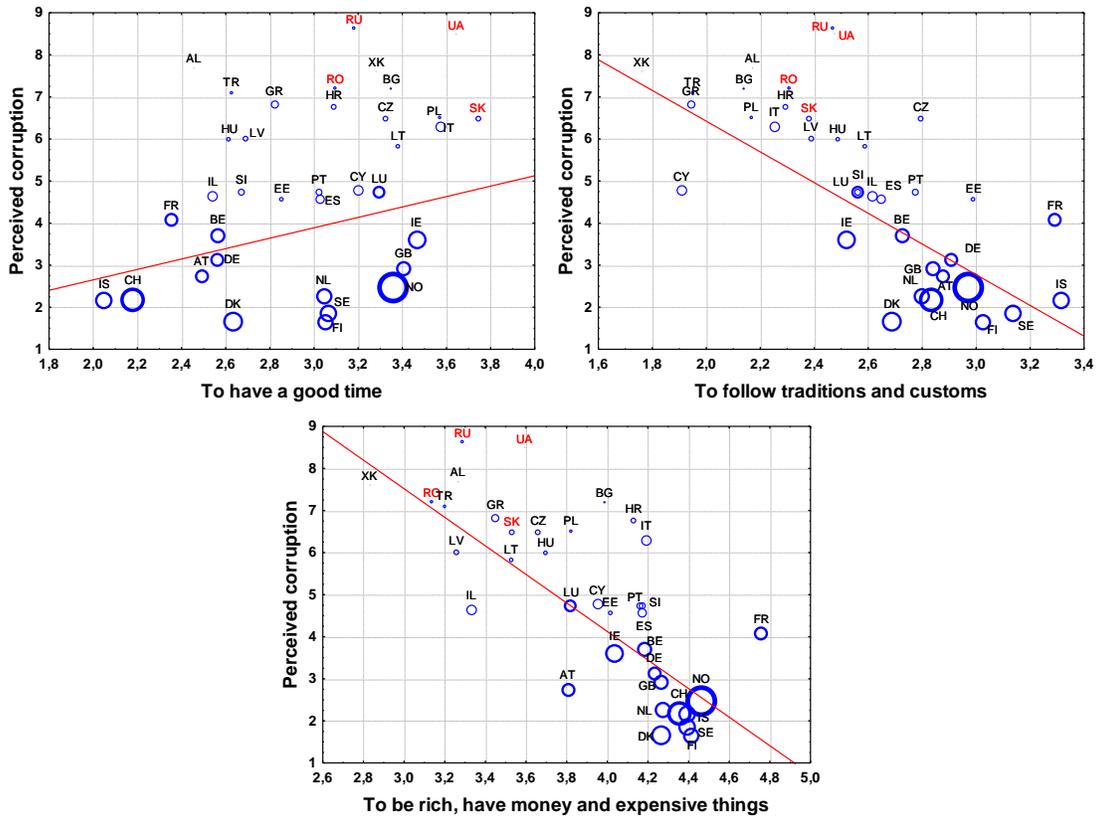

**FIGURE 2: Association between human values and perceived corruption in 36 European countries.**
Scatter plots of the relationship between individual values and perceived corruption for participant countries in the "European Social Survey". Bubble size indicates per capita GDP (the bigger the bubble the wealthier the country). Countries in red have numerous synonyms for the word "guilt".

Table 4 presents two stepwise regression models for predicting corruption perception based on the prevailing division of labor in the production of knowledge. Model 3 shows the indicators of academic productivity that work as significant predictors of perceived corruption when GDP per capita is not included in the model's predictors ($R^2 = 0.867$; $p < 0.001$). Model 2 shows that only one of these predictors proved to be statistically significantly for predicting corruption perception along with GDP per capita ($R^2 = 0.846$; $p < 0.001$). Table 4 shows that the relationships between wealth, corruption and the form of division of labor used in the production of knowledge are intimately entangled and statistically not different.

**Table 4: Multivariate relationship of corruption perception, economic wealth and academic productivity**

| Best predictors | Model 1 | Model 2 | Model 3 |
|---|---|---|---|
| GDP per capita (current US$) | -0.87 | -0.50 (0.12) | |
| S-Share Biochemistry, genetics and molecular biology | | -0.38 (0.10) | -0.79*** (0.11) |
| S-Share Business, management & accounting | | -0.26 (0.08) | -0.43*** (0.11) |
| S-Share chemical engineer | | | -0.35** (0.13) |
| S-share chemistry | | | 0.43*** (0.18) |
| S-Share Earth & planetary sciences | | | -0.30* (0.11) |
| S-Share Materials science | | | -0.83*** (0.22) |
| S-Share Nursing | | | -0.46*** (0.11) |
| S-Share Physics & astronomy | | | 0.43** (0.19) |
| R-Squared | 0.77 | 0.85*** | 0.87*** |

NOTE: Each column reports the beta coefficients from stepwise multiple regression models along with its standard errors reported in parentheses. The multivariate association R-squared for each model is presented at the bottom of the table.
* Correlation is significant at p < 0.05; ** significant at p < 0.01; *** significant at p < 0.001

The scientific areas whose relative productivity correlated with perceived corruption were those identified for rich countries in a worldwide study [36]. Interestingly, the share of publications in neuroscience correlated with corruption perception even after filtering out the effect of GDP per capita. This result might be explained by the fact that countries like Iceland and Hungary showed a high S-share in neuroscience despite having relatively low wealth but have lower corruption perception than what is expected for their GDP per capita (See Figure 3). The relative productivity in business sciences did not correlate with perceived corruption or wealth (Figure 3).

**Table 5: Association between indicators of academic productivity and perceived corruption, guilt synonyms, publications per capita, economic wealth and happiness for participant countries in the ESS**

|  | Perceived corruption | Guilt synonyms | Publication per capita | GDP per capita (current US$) | How happy are you |
|---|---|---|---|---|---|
| Nursing | -0,74*** | -0,27 | 0,25 | 0,77*** | 0,78*** |
| Immunology and microbiology | -0,74*** | -0,24 | 0,17 | 0,74*** | 0,71*** |
| Neuroscience | -0,72*** | -0,11 | 0,49** | 0,67*** | 0,64*** |
| Health professions | -0,70*** | -0,38* | 0,29 | 0,73*** | 0,69*** |
| Psychology | -0,68*** | -0,27 | 0,21 | 0,70*** | 0,73*** |
| Biochemistry | -0,66*** | -0,14 | 0,50** | 0,61*** | 0,62*** |
| Dentistry | -0,59*** | -0,41 | 0,35 | 0,65*** | 0,62*** |
| Medicine | -0,55*** | -0,31 | 0,55*** | 0,62*** | 0,62*** |
| Multidisciplinary | -0,42** | 0,07 | 0,17 | 0,37 | 0,39** |
| Decision science | -0,32 | -0,25 | 0,28 | 0,44 | 0,35* |
| Business, management & accounting | -0,32 | -0,47** | -0,36* | 0,31 | 0,28 |
| Social sciences | -0,29 | -0,41** | -0,49** | 0,25 | 0,28 |
| Arts & humanities | -0,29 | -0,20 | -0,36* | 0,26 | 0,31 |
| Environmental science | -0,29 | -0,40 | -0,27 | 0,18 | 0,26 |
| Earth & planetary sciences | -0,23 | -0,10 | 0,12 | 0,21 | 0,23 |
| Pharmacology | -0,21 | -0,19 | 0,31 | 0,20 | 0,29 |
| Agricultural & biological sciences | -0,18 | -0,48** | -0,32 | 0,06 | 0,16 |
| Veterinary | -0,05 | -0,08 | -0,17 | -0,02 | 0,02 |
| Economics, econometrics & finance | -0,04 | -0,25 | -0,32 | 0,20 | 0,12 |
| Energy | 0,09 | -0,08 | -0,06 | -0,06 | -0,23 |
| Computer science | 0,32 | -0,21 | -0,25 | -0,27 | -0,26 |
| Physics and astronomy | 0,44** | 0,39** | 0,02 | -0,46** | -0,50** |
| Materials science | 0,46** | 0,40** | -0,02 | -0,50** | -0,57*** |
| Engineering | 0,46** | 0,18 | -0,23 | -0,43** | -0,53** |
| Chemical engineer | 0,49** | 0,20 | 0,12 | -0,53** | -0,49** |
| Chemistry | 0,51** | 0,37* | 0,19 | -0,56*** | -0,54*** |
| Mathematics | 0,54*** | 0,35* | -0,03 | -0,46** | -0,49** |

NOTE: The table presents the correlation between the indicators of scientific productivity (as captured by the database of SCImago) and perceived corruption, wealth, synonyms of guilt and happiness for participant countries in the ESS. The correlations are estimated through Spearman's Rho non-parametric correlation coefficient. * Significant correlation at p < 0.05; ** at p < 0.01; *** at p < 0.001.

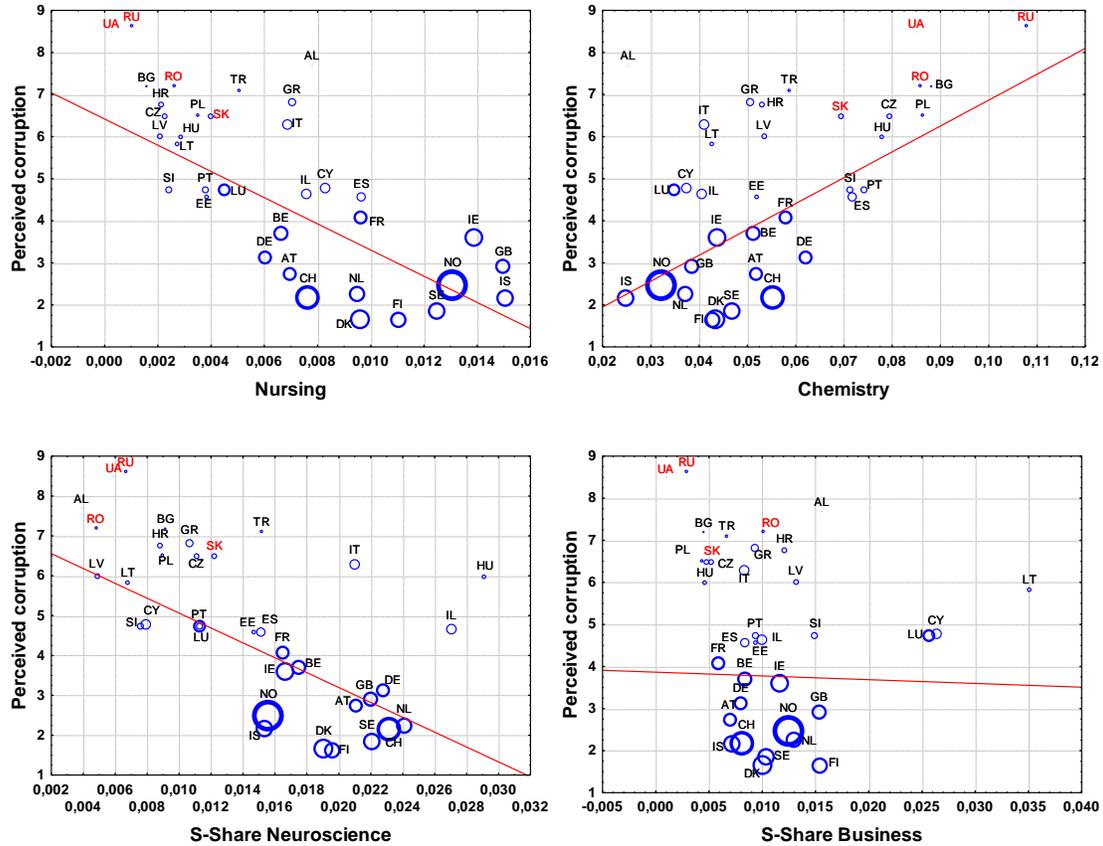

**FIGURE 3: Association between academic productivity and perceived corruption in 36 European countries.** Scatter plots of the relationship between four of the most important indicators of academic productivity in Table V (Nursing, Neuroscience, Business and Chemistry) and perceived corruption. Bubble size indicates GDP per capita for each nation (i.e., the bigger the bubble, the richer the country). Countries in red have numerous synonyms for the word "guilt".

Table 5 presents the non-parametric correlations of scientific productivity with corruption perception, GDP per capita, subjective happiness and the number of recognized synonyms for the word "guilt". Countries with higher levels of perceived corruption by its inhabitants are more productive in "Chemistry", "Physics and astronomy", "Materials sciences" and "Mathematics". These correlations are quite stable in time during the last decade (see Table A1 in appendix). Nations with the lowest levels of perceived corruption are relatively more productive in "Health professions" and "Nursing". Figure 3 depicts the relationship between per capita GDP and scientific productivity in Chemistry and Nursing for participant countries of the ESS.

Cultural values also show significant associations with perceived corruption, the economic wealth and the subjective happiness. Table V shows the non-parametric correlations of perceived corruption, economic wealth, subjective happiness and the number of synonyms for the word "guilt". Figure 3 highlights with colors the cultures based on "guilt" societies in these relationships. It is very striking to see how counties speaking languages with the most synonyms for guilt, such as Russian, Romanian, Ukrainian and Slovakian, are among the poorest and most corrupt in Europe.

DISCUSSION

Our analysis confirmed that most of the covariates to corruption that had been reported in the literature as a result of comparisons between the countries of the world, are also acting among the European countries, except that with taxations, which our data did not captured. This study revealed the existence of a complex web of relationships between the personal values of individuals in different societies with perceived corruption, subjective happiness, economic wealth and the linguistic importance assigned to the concept of guilt. The relationship between these variables is very strong and intimate and difficult to untangle statistically. We have no way to determine the causal link among these variables through their statistical relationships. We have indications, however, that wealth changes faster than corruption perception. Countries like Norway and Russia increased their wealth in the last decades thanks to oil-exports. Their level of perceived corruption, however, has not decreased correspondingly. Historic data from ESS also show very stable outcomes of European surveys on individual values (see appendix). Clearly enough, perceived corruption and the economic performance of a nation are part of a syndrome that includes individual values of economic agents and the way labor in knowledge production is divided in a country. The most likely causal relationship between these variables is mutual reinforcement, where for example corrupt countries are poor, enhance values that are compatible with this situation, fomenting corruption further, and thus have undeveloped scientific and economic ecosystems.

The role of rational knowledge production, best estimated through the way scientific productivity is organized in a country, is a key predictor of economic wealth. The structure of knowledge, however, changes as the scientific system becomes more mature. Countries with incipient scientific establishments vary greatly in the emphasis they place in different areas of knowledge. Middle income countries that develop relatively stronger basic sciences produce faster economic growth compared to countries that do not [50]. In Europe, however, all countries have rather old scientific establishments and have fluent contact among them. Joining the insight gained in this study with previous ones, we can postulate that at least three different evolutionary phases in the relationship of wealth and the structure of scientific system of a country exist. i) Very poor societies have no or very incipient scientific research activities. ii) Societies eventually start to develop a scientific research community. Those nations with incipient scientific systems that give more emphasis to basic sciences are more successful than countries putting their research efforts elsewhere. iii) Countries with well-established scientific communities, start developing science by focusing on novel and more diverse aspects, diluting the relative research efforts made in basic sciences (as shown here). The last two of these phases can be identified in Figure 4.

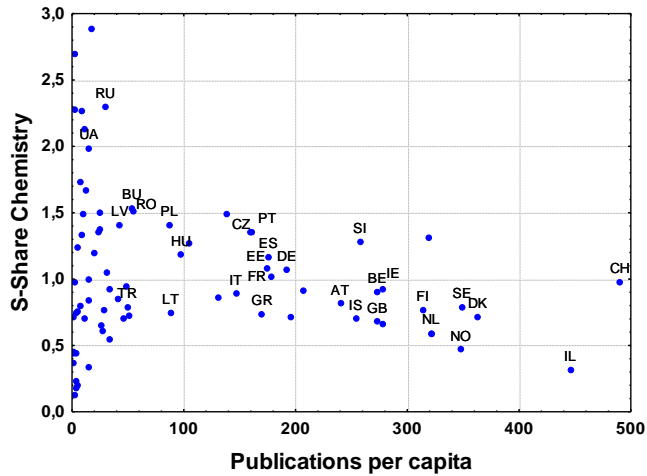

**FIGURE 4: Association between Publications per capita and the share of publications in the area of chemistry in a large selection of countries of the world.** This scatterplot depicts the relationship between the share of publications in chemistry in countries with different overall rates of scientific productivity as measured by publications per capita in a large selection of countries in the world. Only European countries are indicated with their two letter abbreviation.

This explains why corrupt countries in Europe emphasize more basic sciences than less corrupt and wealthier nations. These countries are also the ones with the lowest scientific development in Europe. This interesting correlation is very likely related to the fact that in more developed countries scientific rationality is more dominant, leading to an increasing overall productivity and increased well-being of its inhabitants, through evidence-based public policies [54]. In Europe, more scientifically and more developed nations like Norway, Denmark or Finland have higher scientific productivity in health-related disciplines compared to Russia, Romania, Slovakia and Ukraine. This difference in the division of labor in the production of knowledge is a reflection of differences in overall scientific development. It also shows that for prosperous countries it is important to maintain and develop institutions with scientific research purposes that eventually benefit society by providing the scientific base that is required for creating useful technology in business activities [55].

If one accepts that prevalent social values affect the quality of public institutions, its norms and prosecutors [52], then it is easier to understand the occurrence of corrupt behaviors like nepotism and bribes that are clearly linked with the importance of being rich, or being loyal to friends, that we observed strongly associated with corruption perception, economic wealth and subjective happiness. Most of the human values assessed in the ESS do not differentiate statistically "corrupt" and "non-corrupt" countries. Only those related with Achievement and Security are clearly correlated with a lack of corruption. This is in accordance with previous results cited above that suggested that bureaucracies where corruption is higher are less likely to provide a strong bulwark against infringements on property rights, producing distortions in investment and trade that may reduce the quantity and efficiency of capital investment and foreign technology introduced into the country [56]. This finding is also related with recent observations regarding the role of trust of citizens in public institutions; more specifically that corruption can inhibit economic development by eroding confidence in public institutions.

Previous studies found a strong relationship between corruption, culture and language. Coincidentally, we found that countries with several synonyms for the word "guilt" (five or more) also proved to be the ones with higher levels of perceived corruption, while nations with only one synonym for this word were among the cleanest states. This result might be interpreted as evidence that support the relationship between language and future-oriented economic behaviors that was recently observed [44]. Future works should be extended to other regions of the world (Latin-American and/or Asian nations) in order to evaluate the consistency of our results.

Accepting that corruption is part of a socio-cultural-economic phenomenon forces us to recognize that optimal policies to foment economic growth should be different in poor, corrupt countries with low scientific-technological development than in countries with highly developed industries that have shown their own ability to produce wealth. Our findings corroborate that this difference includes individual, social and cultural values. Thus, policies that do not take into account these differences are doomed to fail. Experimentation at small scales, addressing values and attitudes, together with the implementation of economic policies, should help by improving the success rate of economic policies. This recommendation seems to be especially pertinent for international efforts in triggering economic growth that have failed spectacularly so far, in Haiti, Afghanistan, Iraq and other countries; as well as for nations like Ukraine where such efforts are being initiated.

The most conspicuous finding, however, is that that economic prosperity is the derived state of a nation, whereas corruption and poverty is the original state. Economic evolution in human history went from poverty to prosperity [57], although examples of countries that went through the inverse route also exist, such as the recent history of Zimbabwe and Venezuela [58]. This general trend can be deduced from the fact that the strongest correlations between prevailing individual values and corruption is a negative one for value dimensions related to personal achievements and security. Positive correlations between individual values and corruption were very feeble. That is, countries that educate its citizens to value security and personal achievements are more prone to assemble complex interwoven societies with strong institutions. We propose therefore to define a syndrome of prosperity, rather than one for corruption, and to consider corruption as the primitive state in social evolution, prevalent in societies that value more family ties than abstract law. Evolutionary talking, the emergence of prosperous modern technological societies was based on their rational knowledge and law. Cultures have to evolve in order to allow the emergence of this new society, although cultures change slower than economies [35]. Despite the evident triviality of this insight, economists have focused more in understanding the prevalence of corruption in society rather than focusing on its absence (see the large body of literature mentioned in the introduction). Finely describing the prosperity syndrome might help visionaries in different poor nations to find an appropriate path, adapted to its culture and possibilities, to eventually achieve prosperity.

# 1 Appendix

Table A1: List of participant countries in the six rounds of the ESS

| Participating country | 2002 | 2004 | 2006 | 2008 | 2010 | 2012 |
|---|---|---|---|---|---|---|
| Albania (AL) | No | No | No | No | No | Yes |
| Austria (AT) | Yes | Yes | Yes | No | No | No |
| Belgium (BE) | Yes | Yes | Yes | Yes | Yes | Yes |
| Bulgaria (BU) | No | No | Yes | Yes | Yes | Yes |
| Switzerland (CH) | Yes | Yes | Yes | Yes | Yes | Yes |
| Cyprus (CY) | No | No | Yes | Yes | Yes | Yes |
| Czech Republic (CZ) | Yes | Yes | No | Yes | Yes | Yes |
| Germany (DE) | Yes | Yes | Yes | Yes | Yes | Yes |
| Denmark (DK) | Yes | Yes | Yes | Yes | Yes | Yes |
| Estonia (EE) | No | Yes | Yes | Yes | Yes | Yes |
| Spain (ES) | Yes | Yes | Yes | Yes | Yes | Yes |
| Finland (FI) | Yes | Yes | Yes | Yes | Yes | Yes |
| France (FR) | Yes | Yes | Yes | Yes | Yes | Yes |
| Great Britain (GB) | Yes | Yes | Yes | Yes | Yes | Yes |
| Greece (GR) | Yes | Yes | No | Yes | Yes | No |
| Croatia (HR) | No | No | No | Yes | Yes | No |
| Hungary (HU) | Yes | Yes | Yes | Yes | Yes | Yes |
| Ireland (IE) | Yes | Yes | Yes | Yes | Yes | Yes |
| Iceland (IL) | Yes | No | No | Yes | Yes | Yes |
| Israel (IS)[*] | No | Yes | No | No | No | Yes |
| Italy (IT) | Yes | No | No | No | No | Yes |
| Lithuania (LT) | No | No | No | No | Yes | Yes |
| Luxembourg (LU) | Yes | Yes | No | No | No | No |
| Latvia (LV) | No | No | No | Yes | No | No |
| Netherlands (NL) | Yes | Yes | Yes | Yes | Yes | Yes |
| Norway (NO) | Yes | Yes | Yes | Yes | Yes | Yes |
| Poland (PL) | Yes | Yes | Yes | Yes | Yes | Yes |
| Portugal (PT) | Yes | Yes | Yes | Yes | Yes | Yes |
| Romania (RO) | No | No | No | Yes | No | No |
| Russia (RU) | No | No | Yes | Yes | Yes | Yes |
| Sweden (SE) | Yes | Yes | Yes | Yes | Yes | Yes |
| Slovenia (SI) | Yes | Yes | Yes | Yes | Yes | Yes |
| Slovakia (SK) | No | Yes | Yes | Yes | Yes | Yes |
| Turkey (TR) | No | Yes | No | Yes | No | No |
| Ukraine (UA) | No | Yes | Yes | Yes | Yes | Yes |
| Kosovo (XK) | No | No | No | No | No | Yes |
| Participant countries | 22 | 25 | 23 | 29 | 27 | 29 |

[*] Israel was the only Non-European country included in the European Social Survey

Table A2: Statistical non-parametric correlations between GDP per capita and perceived corruption in participant countries of ESS

|  | 2002 | 2004 | 2006 | 2008 | 2010 | 2012 |
|---|---|---|---|---|---|---|
| Spearman Correlation | -0.830** | -0.872** | -0.905** | -0.923** | -0.893** | -0.902** |

* Correlation is significant at p < 0.05; ** significant at p < 0.01; *** significant at p < 0.001

Table A3: Correlations between scientific productivity and perceived corruption, wealth and subjective happiness in participant countries of the European Social Survey

| S-Share | Year | Perceived corruption | Per capita GDP | Subjective happiness |
|---|---|---|---|---|
| Chemistry | 2002 | -0.80** | -0.87** | -0.78** |
|  | 2004 | -0.71** | -0.82** | -0.71** |
|  | 2006 | -0.93** | -0.85** | -0.86** |
|  | 2008 | -0.77** | -0.69** | -0.66** |
|  | 2010 | -0.68* | -0.63* | -0.56* |
|  | 2012 | -0.55* | -0.55* | -0.58* |
| Materials Science | 2002 | -0.59* | -0.71** | -0.61** |
|  | 2004 | -0.67** | -0.68** | -0.73** |
|  | 2006 | -0.85** | -0.78** | -0.80** |
|  | 2008 | -0.67* | -0.61* | -0.61* |
|  | 2010 | -0.60* | -0.60* | -0.63* |
|  | 2012 | -0.50* | -0.44* | -0.44* |
| Mathematics | 2002 | -0.72** | -0.81** | -0.79** |
|  | 2004 | -0.55* | -0.55* | -0.52* |
|  | 2006 | -0.79** | -0.67** | -0.71** |
|  | 2008 | -0.62* | -0.48* | -0.44* |
|  | 2010 | -0.67** | -0.53* | -0.52* |
|  | 2012 | -0.51* | -0.43* | -0.37* |
| Physics and Astronomy | 2002 | -0.65** | -0.78** | -0.71** |
|  | 2004 | -0.60* | -0.60* | -0.71** |
|  | 2006 | -0.84** | -0.70** | -0.76** |
|  | 2008 | -0.66* | -0.53* | -0.55* |
|  | 2010 | -0.66* | -0.59* | -0.65** |
|  | 2012 | -0.58* | -0.46* | -0.53* |
| Health Professions | 2002 | 0.68** | 0.71** | 0.65** |
|  | 2004 | 0.56* | 0.62* | 0.58* |
|  | 2006 | 0.89** | 0.87** | 0.87** |
|  | 2008 | 0.77** | 0.74** | 0.66** |
|  | 2010 | 0.84** | 0.88** | 0.78** |
|  | 2012 | 0.78** | 0.76** | 0.76** |
| Nursing | 2002 | 0.77** | 0.81** | 0.75** |
|  | 2004 | 0.84** | 0.85** | 0.80** |
|  | 2006 | 0.80** | 0.81** | 0.79** |
|  | 2008 | 0.80** | 0.87** | 0.74** |
|  | 2010 | 0.73** | 0.82** | 0.69** |

|  | 2012 | 0.73** | 0.77** | 0.75** |

* Correlation is significant at p < 0.05; ** significant at p < 0.01; *** significant at p < 0.001

Table A4. The ten basic values in the ESS survey

| VALUE and central goal | Items that measure each value with their ESS labels |
|---|---|
| **POWER** Social status and prestige, control or dominance over people and resources. | 1- It is important to him to be rich. He wants to have a lot of money and expensive things.<br>2- It is important to him to get respect from others. He wants people to do what he says. |
| **ACHIEVEMENT** Personal success through demonstrating competence according to social standards. | 3- It is important to him to show his abilities. He wants people to admire what he does.<br>4- Being very successful is important to him. He hopes people will recognize his achievements. |
| **HEDONISM** Pleasure and sensuous gratification for oneself. | 5- He seeks every chance he can to have fun. It is important to him to do things that give him pleasure.<br>6- Having a good time is important to him. He likes to "spoil" himself. |
| **STIMULATION** Excitement, novelty, and challenge in life. | 7- He likes surprises and is always looking for new things to do. He thinks it is important to do lots of different things in life.<br>8- He looks for adventures and likes to take risks. He wants to have an exciting life. |
| **SELF DIRECTION** Independent thought and action choosing, creating, exploring. | 9- Thinking up new ideas and being creative is important to him. He likes to do things in his own original way.<br>10- It is important to him to make his own decisions about what he does. He likes to be free and not depend on others. |
| **UNIVERSALISM** Understanding, appreciation, tolerance and protection for the welfare of all people and for nature. | 11- He thinks it is important that every person in the world should be treated equally. He believes everyone should have equal opportunities in life.<br>12- It is important to him to listen to people who are different from him. Even when he disagrees with them, he still wants to understand them.<br>13- He strongly believes that people should care for nature. Looking after the environment is important to him. |
| **BENEVOLENCE** Preservation and enhancement of the welfare of people with whom one is in frequent personal contact. | 14- It is very important to him to help the people around him. He wants to care for their well-being.<br>15- It is important to him to be loyal to his friends. He wants to devote himself to people close to him. |
| **TRADITION** Respect, commitment and acceptance of the customs and ideas that one's culture or religion impose on the individual. | 16- It is important to him to be humble and modest. He tries not to draw attention to himself.<br>17- Tradition is important to him. He tries to follow the custom handed down by his religion or his family. |
| **CONFORMITY** Restraint of actions, inclinations, and impulses likely to upset or harm others and violate social expectations or norms. | 18- It is important to him always to behave properly. He wants to avoid doing anything people would say is wrong.<br>19- He believes that people should do what they are told. He thinks people should follow rules at all times, even when no-one is watching. |
| **SECURITY** Safety, harmony and stability of society, of relationships, and of self. | 20- It is important to him to live in secure surroundings. He avoids anything that might endanger his safety.<br>21- It is important to him that the government ensures his safety against all threats. He wants the state to be strong so it can defend its citizens. |